# Impact of the circuit layout on the charge collection in a monolithic pixel sensor

C. Lemoine,[a,b,1] R. Ballabriga,[a] E. Buschmann,[c] M. Campbell,[a]
R. Casanova Mohr,[d] D. Dannheim,[a] J. Dilg,[e] A. Dorda,[a] F. King,[e] O. Feyens,[e]
P. Gadow,[a] I.M. Gregor,[e] K. Hansen,[e] Y. He,[e] L. Huth,[e] I. Kremastiotis,[a]
S. Maffessanti,[e] L. Mendes,[e] Y. Otarid,[a] C. Reckleben,[e] S. Rettie,[a]
M.A. del Rio Viera,[e] S. Ruiz Daza,[e] J. Schlaadt,[e] A. Simancas,[e] W. Snoeys,[a]
S. Spannagel,[e] T. Vanat,[e] A. Velyka,[e] G. Vignola,[e] H. Wennlöf[e,2]

[a]*CERN,*
  *Esplanade des Particules 1, Geneva, Switzerland*
[b]*IPHC, Université de Strasbourg,*
  *23 rue du Loess, Strasbourg, France*
[c]*Brookhaven National Laboratory (BNL),*
  *New York 11973-5000, Upton, USA*
[d]*Institut de Física d'Altes Energies (IFAE),*
  *Edifici CN, UAB campus, 08193 Bellaterra (Barcelona), Spain*
[e]*Deutsches Elektronen-Synchrotron DESY,*
  *Notkestr. 85, 22607 Hamburg, Germany*

  E-mail: corentin.lemoine@cern.ch

Abstract: CERN's strategic R&D programme on technologies for future experiments recently started investigating the TPSCo 65nm ISC CMOS imaging process for monolithic active pixels sensors for application in high energy physics. In collaboration with the ALICE experiment and other institutes, several prototypes demonstrated excellent performance, qualifying the technology. The Hybrid-to-Monolithic (H2M), a new test-chip produced in the same process but with a larger pixel pitch than previous prototypes, exhibits an unexpected asymmetric efficiency pattern.

This contribution describes a simulation procedure combining TCAD, Monte Carlo and circuit simulations to model and understand this effect. It proved able to reproduce measurement results and attribute the asymmetric efficiency drop to a slow charge collection due to low amplitude potential wells created by the circuitry layout and impacting efficiency via ballistic deficit.

---

[1]Corresponding author.
[2]Now at Nikhef, Amsterdam.

# 1 Introduction

CERN's strategic R&D programme on technologies for future experiments pursues the development of future high-precision Monolithic Active Pixel Sensors (MAPS) for high energy physics (HEP). The TPSCo 65nm ISC (Image Sensor CMOS) process is the first candidate technology for this purpose and thanks to several prototypes such as the APTS[1] and the DPTS[2] it is now qualified for HEP. These prototypes demonstrated that the full chain of sensor, front end and readout electronics achieves full detection efficiency for minimum ionizing particles at an acceptable fake hit rate before and after irradiation up to $10^{15}$ 1MeV $n_{eq}$ cm$^{-2}$.

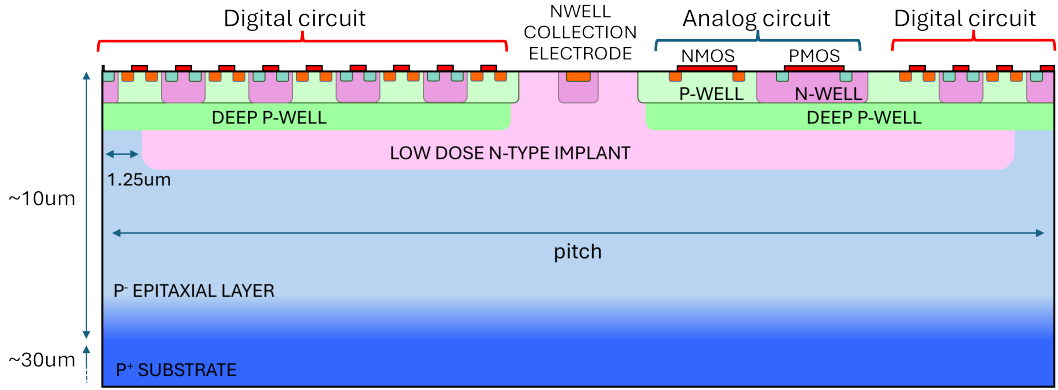

**Figure 1**: Cross section of a typical sensor in the candidate technology. Not to scale.

Figure 1 shows the cross section of a typical sensor in this process with a n-type collection electrode in a low doped p-type epitaxial layer. As introduced in [3], the so-called *modified with gap* layout is used to improve the performance of the sensor by adding an additional low dose deep n-type implant in the pixel, moving the junction deeper in the sensor thus allowing larger depletion and higher vertical electric field. The gap in this deep implant increases the lateral electric field at the pixel edge, concentrating most of the charge in a single pixel and leading to a higher signal to noise ratio. The in-pixel circuitry surrounds the collection electrode, with a deep p-well shielding the n-wells from the collection volume to avoid parasitic charge collection.

**The Hybrid to Monolithic chip** The Hybrid to Monolithic (H2M)[4] is a test-chip designed to port an existing hybrid sensor architecture into a monolithic chip. It features a pixel pitch of 35 μm, larger than any other prototypes tested so far in this technology. Each pixel is equipped with a charge sensitive amplifier (CSA) with Krummenacher feedback [5] which biasing current sets the slope of the return to baseline following a hit. The analogue front end has a relatively short dead time: at nominal settings, simulation predict a return to baseline in few hundreds of ns for an input charge of 600 e$^-$ corresponding to a minimum ionizing particle.

The H2M is fully operational, but in some conditions specified in [4] measurement results exhibit an unexpectedly asymmetric efficiency drop in some parts of the pixel, suggesting some impact of the circuit layout on the sensor performance. So far, no such effect was observed on any other prototype in the same technology, all with a pixel pitch of 25 μm or less.

This paper describes a simulation procedure able to reproduce the asymmetric efficiency pattern measured on the H2M chip and attributed to the layout of the n-wells in the pixel matrix. Details on the chip and measurement results can be found in [4].

## 2 Simulation procedure

The simulation procedure combines a Technology Computer Aided Design (TCAD) tool to compute the electric field inside the sensor with a Monte Carlo tool to simulate charge deposition and transport. This approach allows the complex electric field inherent to the process to accurately be taken into account while being able to simulate a large number of events to model stochastic effects, as they appear in a real particle interaction. It gives access to the collected charge as a function of time, or equivalently to the current pulse at the input of the front end, for different events produced by a physics simulator. Current pulses are then used in a circuit simulation of the front-end electronics.

### 2.1 TCAD simulation

The TCAD simulation of the sensor is performed with Sentaurus [6] from Synopsys by first generating a 3D finite element model of the sensor containing the doping concentration information before numerically solving the electrical behaviour equations inside the device.

The simulated structure includes the n-wells and p-wells, the deep p-wells, the deep n-type implant, the epitaxial layer and a few micrometers of the substrate. The layout of the n-wells is visible in the right part of figure 2 and in the cross section of figure 1. The transistors (shown for illustration in figure 1) and the back end of line (not shown) are not included in the simulation.

The doping concentrations used in the simulation are based on information communicated by the manufacturer and prevent publication of the electric field. We can anyway mention that due to the large pitch, a very low lateral electric field can be observed in the deep n-type implant in the region far from the collection electrode and the gap in the deep n-type implant. This region with very low electric field makes the structure sensitive to small local perturbations of this electric field. As a consequence, below the n-wells of the circuitry, low amplitude potential wells can be observed in the deep n-type implant, with a shape dependent on the layout.

Not only the electric field but also the charge propagation can fully be simulated within TCAD, for example for a particle traversing the sensor with a perpendicular incidence and depositing charge uniformly along its path in the sensor. The impact of the n-wells on the charge collection can be demonstrated by performing such a simulation with or without n-wells and comparing the results: figure 2 shows a significantly slower charge collection with the addition of n-wells. The total collected charge after a long enough integration time converges to a similar value. The 10% difference can be explained by the fact that a fraction of the charge is deposited within an n-well which is part of the circuitry: in the case with n-wells, the charge deposited in approximately the first micrometer gets collected by said n-well and not by the collection electrode. This is confirmed in the TCAD simulation where $65\,e^-$ are collected by the n-well from the circuit. This happens immediately after the charge deposition and not during the charge transport.

The larger collection time when including the circuitry can be explained by the presence of the low amplitude potential wells below the n-wells that create a weak barrier to the lateral movement of charges. This barrier is enough to significantly slow down charge collection but not to completely prevent it by permanently trapping the charge in the potential well.

TCAD simulations are however computationally intensive, and a Monte Carlo approach is necessary to gather enough statistics to reproduce the asymmetric efficiency pattern of the H2M.

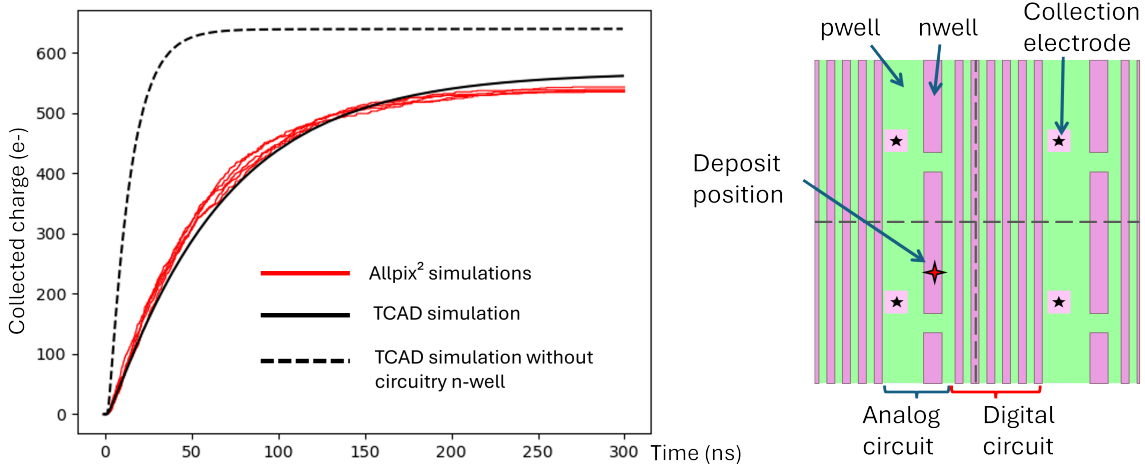

**Figure 2**: (left) Comparison of a signal generated by a TCAD and a Monte Carlo simulation for a 63 e$^-$/µm (arbitrary value) deposition over the full depth. The position of the deposit is indicated in the right part. (right) Simplified layout of the H2M for 2x2 pixels.

## 2.2 Monte Carlo simulation

The charge deposition and propagation inside the sensor are simulated with the Allpix Squared framework [7]. It allows simulation of a physics event with Geant4 [8] to determine the energy deposition in a sensor and computation of the movement of the generated charge carriers within a static electric field imported from TCAD. Charge generation parameters were chosen to reproduce a set-up similar to the H2M test beam, with a perpendicular incident 5 GeV electron beam. Charges are then propagated independently one from another with drift based on the electric field computed in TCAD and diffusion modelled as a 3D gaussian step where the mobility is taken into account via the Masetti-Canali model and recombination is implemented with the SHR-Auger model [9]. To speed up simulations, only electrons are propagated and their charge is only taken into account for the signal when they reach the collection electrode, which is a good approximation for the small collection electrode sensors treated here.

Good agreement between the Monte Carlo in Allpix Squared and a fully self-consistent TCAD simulation including n-wells of the circuitry is demonstrated in figure 2 where the results from several Monte Carlo simulations (to consider event-to-event random variation) are compared to the same simulation in TCAD. This confirms that Allpix Squared can be used for the problem under consideration to replace TCAD by a much faster solution with minimum impact on results.

## 2.3 Circuit simulation

The TCAD and Monte Carlo simulations reproduce only the behaviour of the sensor but do not integrate the front-end electronics. The front end of the H2M is non-linear, as imposed by the nature of the Krummenacher feedback where the discharge rate is almost independent of the input charge. In addition, figure 2 shows that the collection time is of the same order as the front-end response (see section 1). To take the impact of the collection time on the pulse amplitude into account, it is thus required to model the front-end response as a part of the simulation procedure. Therefore, signals generated from Monte Carlo simulation are exported in the form of a current pulse that is injected in a specialized circuit simulator, Spectre [10].

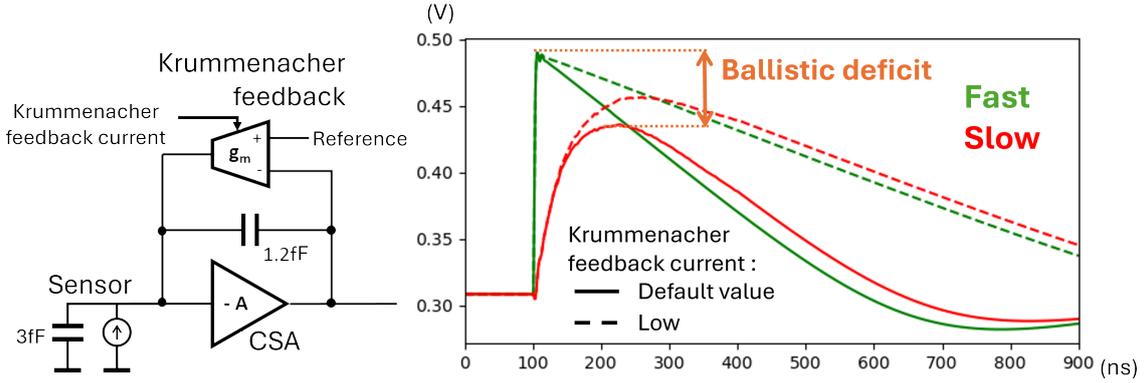

**Figure 3**: (left) Simulated front-end for the H2M chip and (right) simulated output of the CSA for the same input charge (1600e-) but different collection times.

In order to simulate a large number of pulses, only the part of the front-end shown on the left in figure 3 is modeled with a noiseless schematic simulation.

The right plot of figure 3 shows the simulated voltage at the output of the CSA for two inputs both corresponding to the collection of the same amount of charge (1600 e$^-$) but one with a fast collection in the order of ns (green) and one with a slow charge collection in the order of 100 ns (red). These two events were selected from a simulation of the H2M under illumination with X-ray from a $^{55}$Fe radioactive source. The fast event is very close to the expected response, with a fast rise time after the hit followed by a slower return to baseline at constant rate. Due to the charge collection being slow compared to the front-end return to zero, the slow event exhibits a partial return to baseline before complete charge integration (known as ballistic deficit) leading to an underestimation of the collected charge by 30% and confirming the importance of considering this effect with a circuit simulation. Figure 3 also illustrates the impact of the Krummenacher feedback current: by lowering it the return to baseline is slower, thus reducing the impact of ballistic deficit.

## 2.4 Overall approach

After Spectre simulations, the peak voltage at the output of the CSA is extracted and post-processed by a python script to add a 33 e$^-$ RMS gaussian noise (based on measurements) and compare it with a threshold to determine if a hit was registered. Integrating this with the Monte Carlo simulation truth from Allpix Squared, where the track position is smeared with a 3 μm gaussian to reproduce the telescope tracking resolution, allows an in-pixel efficiency map to be built similar to the one produced from a test beam experiment.

## 3 Results

Figure 4 shows the in-pixel efficiency map obtained from the simulation procedure described in section 2 and compares it with the measured one from [4]. Good qualitative matching is achieved, correctly reproducing both the efficiency value and its asymmetric pattern.

Measurements results in [4] also demonstrate that reducing the Krummenacher feedback current, slowing down the return to zero of the front-end, increases efficiency, confirming experimentally that the observed effect is indeed related to ballistic deficit caused by slow charge collection.

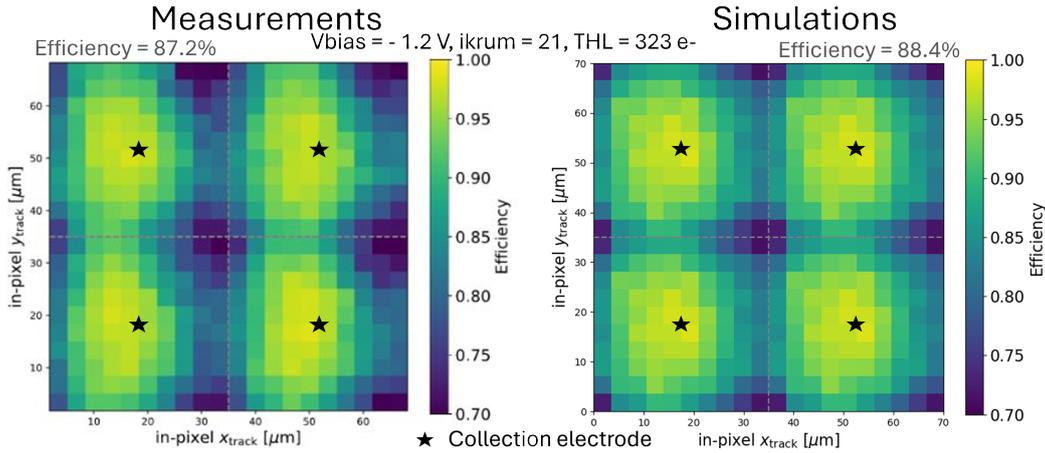

**Figure 4**: Comparison of the in-pixel efficiency map obtained from measurements[4] and from simulation. Results are shown as a stacked 2×2 pixels group separated with a dashed grey line. For the simulation, the plotted data are the same for all 4 pixels.

## 4 Conclusion

This contribution presented a simulation procedure combining TCAD, Monte Carlo and circuit simulations to model complex effects leading to an asymmetric efficiency pattern in the H2M. These effects were attributed to low amplitude potential wells underneath the pixel circuit n-wells.

TCAD simulations showed that, in the very low lateral electric field region inherent to large pitch in this process, small potential wells could cause a clear charge collection slow down. The Monte Carlo approach allowed the simulation of a large number of events mimicking a test beam, and circuit simulations proved crucial to model the ballistic deficit caused by slow signals on the fast return to zero front-end. Combining all steps allowed the production of an in-pixel efficiency map in agreement with measurements.

The large pixel pitch associated with the specific n-well layout and the fast return to baseline front-end explains why similar behaviour was not observed in any other prototypes using the same process. The simulation procedure now enables an evaluation of the impact of the circuit layout on charge collection at the design stage which is crucial for future large pitch and fast front end sensors. Similar simulations confirmed that this effect becomes negligible for smaller pixel pitches, corresponding to the earlier experimental observations.